\documentclass[3p,times]{elsarticle}

\usepackage{ecrc}

\usepackage{sidecap}
\usepackage{ulem}
\usepackage{epsfig}
\usepackage{amsmath,amssymb,amsthm}
\usepackage{graphicx}
\usepackage{bm}
\usepackage{color}

\volume{00}

\firstpage{1}

\journalname{Nuclear Physics A}

\runauth{}

\jid{nupha}

\usepackage{amssymb}

\usepackage[figuresright]{rotating}

\begin{document}
\newcommand{\vAi}{{\cal A}_{i_1\cdots i_n}} 
\newcommand{\vAim}{{\cal A}_{i_1\cdots i_{n-1}}} 
\newcommand{\vAbi}{\bar{\cal A}^{i_1\cdots i_n}}
\newcommand{\vAbim}{\bar{\cal A}^{i_1\cdots i_{n-1}}}
\newcommand{\htS}{\hat{S}} 
\newcommand{\htR}{\hat{R}}
\newcommand{\htB}{\hat{B}} 
\newcommand{\htD}{\hat{D}}
\newcommand{\htV}{\hat{V}} 
\newcommand{\cT}{{\cal T}} 
\newcommand{\cM}{{\cal M}} 
\newcommand{\cMs}{{\cal M}^*}
\newcommand{\vk}{\vec{\mathbf{k}}}
\newcommand{\bk}{\bm{k}}
\newcommand{\kt}{\bm{k}_\perp}
\newcommand{\kp}{k_\perp}
\newcommand{\km}{k_\mathrm{max}}
\newcommand{\vl}{\vec{\mathbf{l}}}
\newcommand{\bl}{\bm{l}}
\newcommand{\bK}{\bm{K}} 
\newcommand{\bb}{\bm{b}} 
\newcommand{\qm}{q_\mathrm{max}}
\newcommand{\vp}{\vec{\mathbf{p}}}
\newcommand{\bp}{\bm{p}} 
\newcommand{\vq}{\vec{\mathbf{q}}}
\newcommand{\bq}{\bm{q}} 
\newcommand{\qt}{\bm{q}_\perp}
\newcommand{\qp}{q_\perp}
\newcommand{\bQ}{\bm{Q}}
\newcommand{\vx}{\vec{\mathbf{x}}}
\newcommand{\bx}{\bm{x}}
\newcommand{\tr}{{{\rm Tr\,}}} 
\newcommand{\bc}{\textcolor{blue}}

\newcommand{\beq}{\begin{equation}}
\newcommand{\eeq}[1]{\label{#1} \end{equation}} 
\newcommand{\ee}{\end{equation}}
\newcommand{\bea}{\begin{eqnarray}} 
\newcommand{\eea}{\end{eqnarray}}
\newcommand{\beqar}{\begin{eqnarray}} 
\newcommand{\eeqar}[1]{\label{#1}\end{eqnarray}}
 
\newcommand{\half}{{\textstyle\frac{1}{2}}} 
\newcommand{\ben}{\begin{enumerate}} 
\newcommand{\een}{\end{enumerate}}
\newcommand{\bit}{\begin{itemize}} 
\newcommand{\eit}{\end{itemize}}
\newcommand{\ec}{\end{center}}
\newcommand{\bra}[1]{\langle {#1}|}
\newcommand{\ket}[1]{|{#1}\rangle}
\newcommand{\norm}[2]{\langle{#1}|{#2}\rangle}
\newcommand{\brac}[3]{\langle{#1}|{#2}|{#3}\rangle} 
\newcommand{\hilb}{{\cal H}} 
\newcommand{\pleft}{\stackrel{\leftarrow}{\partial}}
\newcommand{\pright}{\stackrel{\rightarrow}{\partial}}
\begin{frontmatter}



\dochead{}

\title{Heavy flavor suppression in a dynamical QCD medium with finite magnetic mass}


\author{Magdalena Djordjevic}

\address{Institute of Physics Belgrade, University of Belgrade, Pregrevica 118, 11080 Belgrade, Serbia}

\begin{abstract}
Reliable predictions for jet quenching in ultra-relativistic heavy ion collisions require accurate computation of radiative energy loss. While all available energy loss calculations assume zero magnetic mass, in accordance with the one-loop perturbative calculations,  different non-perturbative approaches report a non-zero magnetic mass at RHIC and LHC. We generalized the dynamical energy loss formalism, to consistently include a possibility for existence of non-zero magnetic screening. We show that this generalization indicates a fundamental constraint on electric to magnetic mass ratio, which appears to be supported by lattice QCD simulations. Jet suppression patterns, obtained from this newly developed generalization, show reasonable agreement with the available RHIC and LHC measurements. 
\end{abstract}


\end{frontmatter}


\section{Introduction} 

Jet suppression measurements at RHIC and LHC, and their 
comparison with theoretical predictions, provide a powerful tool for mapping 
the properties of a QCD medium created in ultra-relativistic heavy ion 
collisions~\cite{Brambilla}. This suppression results 
from the energy loss of high energy partons moving through the 
plasma~\cite{suppression}. Therefore, accurate 
calculations of jet energy loss mechanisms are essential for the reliable 
predictions of jet suppression. 

In~\cite{MD_PRC,DH_PRL}, we developed a theoretical formalism for the 
calculation of radiative energy loss in realistic finite size {\it dynamical} 
QCD medium (see also a viewpoint~\cite{Gyulassy_viewpoint}), which abolished a 
static approximation used in previous models (see e.g.~\cite{GLV,Gyulassy_Wang,Wiedemann}). In this proceedings, we will first provide a brief review of the developed dynamical energy loss formalism. We will then show that in the dynamical medium there are comparable contributions to the energy loss due to both longitudinally (electric) and transversely (magnetic) polarized gluons. Since we will see that the magnetic contribution is significant, we will next discuss how to introduce finite magnetic mass into our formalism. Finally, we will incorporate dynamical energy loss formalism and finite magnetic mass into numerical procedure, which we will use to obtain suppression predictions for RHIC and LHC experiments. Note that only the main results are presented here. For a more detailed version see~\cite{MD_JPG,MD_PLB,MD_PRC2}, and references therein.

\section{Radiative energy loss - electric and magnetic contributions}

In~\cite{MD_PRC,DH_PRL} we calculated the radiative jet energy loss in a finite size dynamical QCD medium. The formalism takes into account that the medium constituents are in reality dynamical, that is they are moving particles, and that the medium has finite size. This dynamical energy loss formalism abolished the previously widely used static approximation. The main result of the formalism is presented by the  Eq.~(\ref{DeltaEDyn}), which shows the expression for the dynamical energy loss. 
%
\beqar
\frac{\Delta E_{\mathrm{rad}}}{E} 
&=& \frac{C_R \alpha_s}{\pi}\,\frac{L}{\lambda_\mathrm{dyn}}  
    \int dx \,\frac{d^2k}{\pi} \,\frac{d^2q}{\pi} \, v(\bq)
        \left(1-\frac{\sin{\frac{(\bk{+}\bq)^2+\chi}{x E^+} \, L}} 
    {\frac{(\bk{+}\bq)^2+\chi}{x E^+}\, L}\right) \,
    2 \, \frac{(\bk{+}\bq)}{(\bk{+}\bq)^2+\chi}
    \left(\frac{(\bk{+}\bq)}{(\bk{+}\bq)^2+\chi}
    - \frac{\bk}{\bk^2+\chi}
    \right) .
\eeqar{DeltaEDyn}
%
Here  $v(\bq)=\frac{\mu_E^2}{\bq^2 (\bq^2+\mu_E^2)}$ is the effective 
crossection in dynamical QCD medium, $\mu_E$ is Debye screening mass, 
$\lambda_\mathrm{dyn}^{-1} \equiv C_2(G) \alpha_s T = 3 \alpha_s T$ 
is defined as ``dynamical mean free path'', 
$\alpha_s = \frac{g^2}{4 \pi}$ is coupling constant and $C_R=\frac{4}{3}$. 
$m_g=\mu_E/\sqrt 2$ is the effective mass for gluons with hard momenta 
$k\gtrsim T$, and $\chi\equiv M^2 x^2 + m_g^2$ where $x$ is the longitudinal 
momentum fraction of the heavy quark carried away by the emitted gluon. We 
assume constant coupling $g$.

Based on this expression, we can compute and compare the energy loss in the static and the dynamic case (see~\cite{DH_PRL}), showing that in dynamical QCD medium the energy loss is significantly larger compared to the static case. Consequently, the question is what is the origin of such energy loss increase in the dynamical QCD medium? To address this question, it is useful to separate the energy loss to the magnetic and the electric contribution (for more details see~\cite{MD_JPG}). Note that the electric contribution comes from the longitudinally polarized gluon exchanges, while the magnetic contribution comes from the transversely polarized gluon exchange. In~\cite{MD_JPG} we showed that, in the dynamical case, there are both electric and magnetic contributions to the energy loss; on the contrary, in the static medium, there is only the electric contribution. 

\begin{figure}[ht]
\vspace*{4.cm} 
\hspace*{2cm}\includegraphics{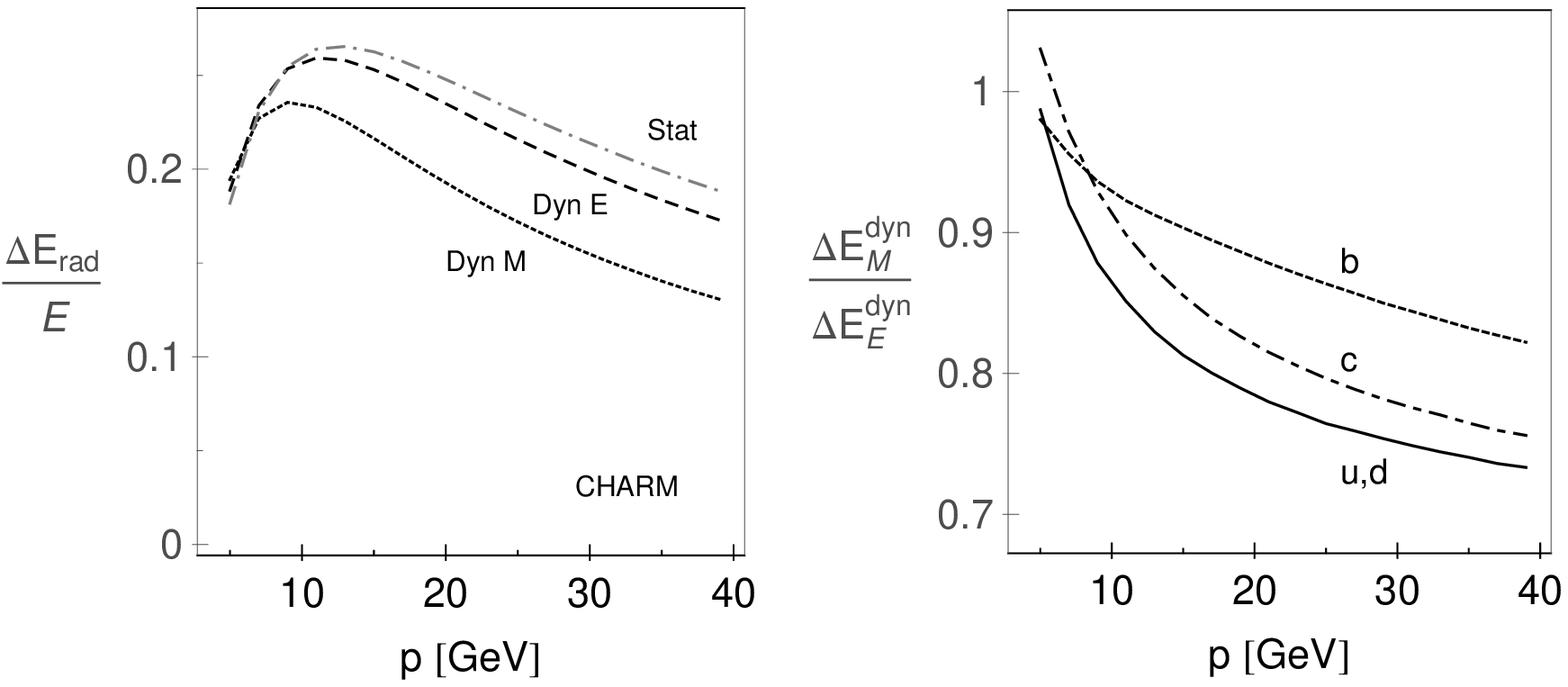}
\caption{{\it Left panel: }Charm quark fractional radiative energy loss as a 
function of 
momentum for an assumed path length $L=5$\,fm at RHIC (medium temperature 
$T=225$\,MeV). Dotted and dashed curves correspond, respectively, to magnetic and electric contributions to the energy loss in a 
dynamical QCD medium, while dot-dashed curve corresponds to the energy loss 
in a static QCD medium (see~\cite{DG_Ind}). {\it Right panel: } Ratio of magnetic and electric contributions to the radiative energy 
loss in a finite size dynamical QCD medium for light, charm and bottom quark 
(full, dot-dashed and dashed curve, respectively) at RHIC. }
\label{Magn_vs_Elec}
\end{figure}

Since both electric and magnetic contributions are present in the dynamical QCD medium, in Fig.~\ref{Magn_vs_Elec} we numerically investigate relative importance of these two contributions. We see that the magnitude of the magnetic and electric contributions are comparable to each other, and they are moreover comparable to the static energy loss. Therefore, we can conclude that the reason for a significant increase of the radiative energy loss in dynamical medium is due to existence of additional magnetic contribution.

\section{Magnetic mass and energy loss}

In the previous section, we showed that magnetic contribution to the dynamical energy loss is important, which opens a question on how this contribution depends on the value of magnetic mass. That is, HTL preturbative QCD, which we used to develop the energy loss formalism, assumes that magnetic mass is equal to zero. On the other hand, different non-perturbative approaches (see e.g.~\cite{Maezawa}) suggest a non-zero magnetic mass at RHIC and LHC. Since magnetic contribution to the energy loss depends on magnetic mass, two questions emerge:  First, can magnetic mass be consistently included in the dynamical energy loss calculations, and if yes, how this inclusion would modify the energy loss results?

To address the above questions, we generalized the dynamical energy loss formalism to the case of finite magnetic mass. From our analysis in~\cite{MD_PLB} follows that finite magnetic mass modifies only the effective crossection in Eq.~(\ref{DeltaEDyn}) to the following expression: 
\beqar
v(\bq) =  \frac{\mu_E^2-\mu_M^2}{(\bq^2+\mu_E^2)(\bq^2+\mu_M^2)}
\eeqar{vqmagn}

Furthermore, it is interesting that inclusion of the finite magnetic mass leads to an important physical constraint on the magnetic mass value. From Eq.~(\ref{vqmagn}) (see also~\cite{MD_PLB}), it follows that, if magnetic mass is larger than electric mass, the quark jet would, overall, start to gain (instead of lose) energy in this type of plasma. Such energy gain would be in an apparent violation of the second law of thermodynamics, since it would involve a transfer of the energy of disordered motion of the medium constituents to the ordered motion of the jet. From this follows a simple constraint that it is not possible to create a plasma with magnetic mass larger than electric. It is interesting that this simple constraint is actually in an agreement with the various non-perturbative approaches, which suggest that, at RHIC and LHC, $0.4<\mu_M/\mu_E<0.6$ (see~\cite{MD_PLB} and references therein).

\begin{figure*}
\vspace*{4.cm} 
\hspace*{2cm}\includegraphics{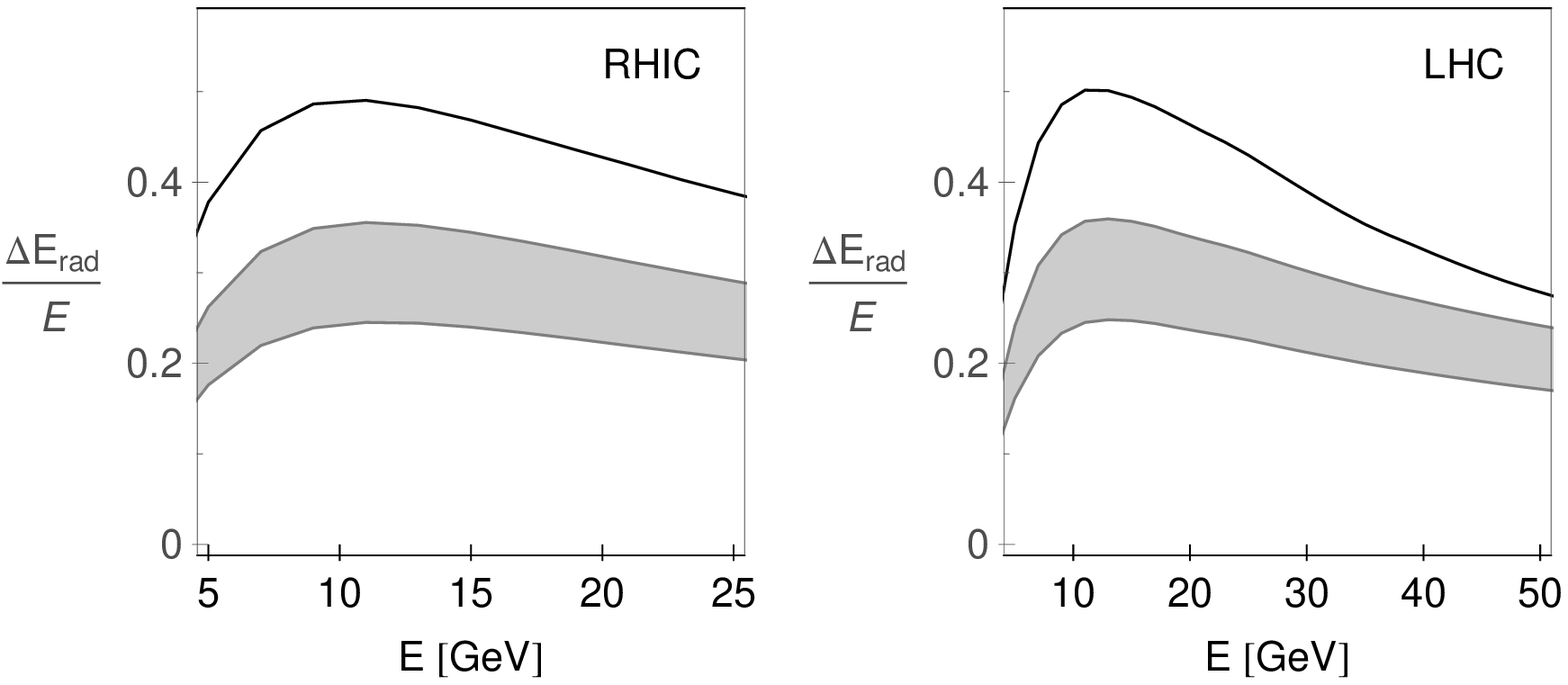}
\caption{Charm quark fractional radiative energy loss as a function of momentum 
for path length of $L=5$\,fm. Full curve corresponds to the case when magnetic mass is zero.  Gray band corresponds to the energy loss when magnetic mass is non-zero (i.e.  $0.4 < \mu_M/\mu_E < 0.6$). Upper (lower) boundary of the band corresponds to the case $\mu_M/\mu_E=0.4$ 
($\mu_M/\mu_E =0.6$).
For left panel, we assume a medium of temperature $T=225$\,MeV and constant coupling $\alpha_S=0.3$ (``RHIC conditions''). For right panel, we assume a medium of temperature $T=290$\,MeV and constant coupling $\alpha_S=0.25$~\cite{Betz} (``LHC conditions''). }
\label{MagnDep}
\end{figure*}

By using Eq.~\ref{vqmagn}, in Fig.~\ref{MagnDep} we numerically explore what is the effect of introducing finite magnetic mass into energy loss calculations.  We see that the introduction of the finite magnetic mass reduces the energy loss for 25\% to 50\% compared to the zero magnetic mass case.

\section{Jet suppression at RHIC and LHC}
 
Based on the developed dynamical energy loss for both zero magnetic value and for finite magnetic value, we now calculate suppression patterns for RHIC and LHC experiments (for more details, see~\cite{MD_PRC2}. In addition to taking into account the dynamical effect and the finite magnetic mass, our numerical procedure also includes multigluon~\cite{GLV_suppress} and pathlength fluctuations~\cite{WHDG}. Multi-gluon fluctuations address the fact that energy loss is a distribution, while path length fluctuations address the fact that particles travel different paths in the medium.

\begin{figure}
\vspace*{4.cm} 
\hspace*{1cm}\includegraphics{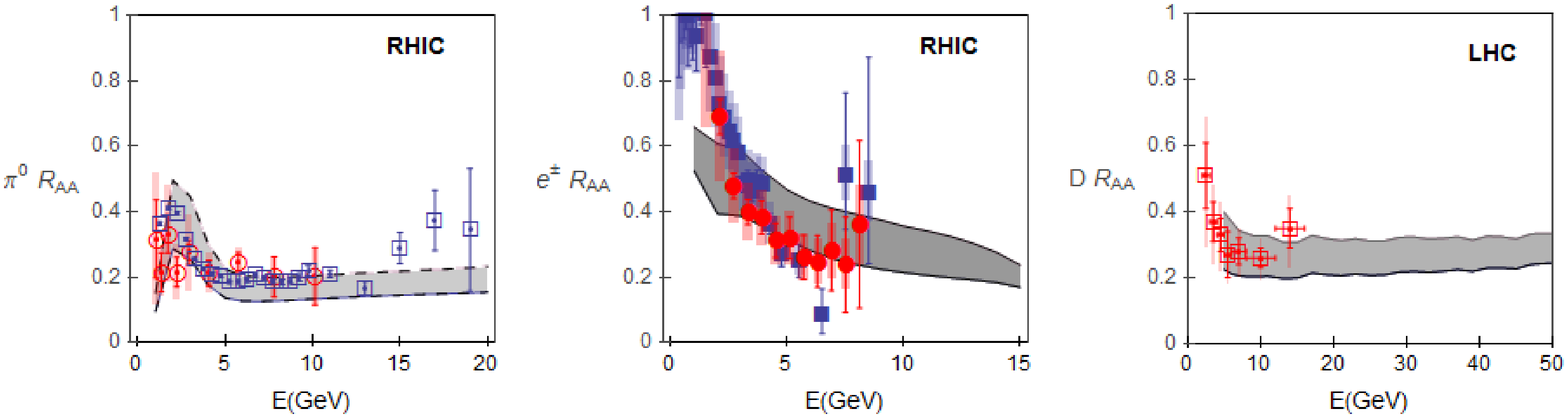}
\caption{(Color online) The left panel shows the comparison of the pion 
suppression predictions with $\pi^0$ PHENIX~\cite{PHENIX_pi0} (shown in blue) 
and STAR~\cite{STAR_pi0} (shown in red) experimental data from 200 GeV Au+Au 
collisions at RHIC. The middle panel shows the comparison of the single 
electron suppression predictions with the non-photonic single electron data 
from PHENIX~\cite{PHENIX_e} (shown in blue) and STAR~\cite{STAR_e} (shown in 
red) at 200 GeV Au+Au collisions. The right panel shows the comparison of the D meson suppression predictions with the available ALICE~\cite{ALICE} data for 2.76TeV Pb+Pb collisions at LHC. The suppression predictions are obtained by using the 
path length distributions from~\cite{WHDG}. We assume constant coupling $\alpha_S=0.3$, and temperature $T=225$~GeV, as representative of RHIC conditions. For LHC conditions, we assume constant coupling $\alpha_S=0.25$, and temperature $T=290$~GeV~\cite{Betz}. On each panel, the gray region 
corresponds to the case when $\mu_M \geq 0$ (i.e.  $0 < \mu_M/\mu_E < 0.6$), 
where the lower boundary corresponds to $\mu_M/\mu_E=0$ and the upper boundary 
corresponds to $\mu_M/\mu_E =0.6$.}
\label{jetsuppression}
\end{figure}

In Fig.~\ref{jetsuppression}, we compare our suppression predictions with available experimental data from RHIC and LHC experiments. Left panel correspond to pions, while middle panel correspond to single electrons from RHIC 200 GeV Au+Au collisions. The right panel shows D meson suppression predictions for 2.76TeV Pb+Pb collisions at LHC. In all three panels, we see a reasonable agreement between our predictions and available experimental data, which is moreover robust with the introduction of the finite magnetic mass value. We can, therefore, conclude that introduction of the dynamical medium significantly improves the agreement between the predictions and experimental data.

\section{Summary}
In this proceedings, we first briefly reviewed the dynamical energy loss formalism, which abolishes the approximation of static medium.
Since we showed that the magnetic contribution to the energy loss is important, we consequently generalized the dynamical
energy loss to include a possibility of finite magnetic mass. This generalization suggests a simple constraint on electric to magnetic mass ratio. We furthermore incorporated  the dynamical formalism, together with the finite magnetic mass, into a numerical procedure for suppression predictions. The predictions for RHIC and LHC indicate a reasonable agreement of the dynamical energy loss formalism with  the available data.

\medskip
{\em Acknowledgments:} 
This work is supported by Marie Curie International Reintegration Grant 
within the $7^{th}$ European Community Framework Programme 
(PIRG08-GA-2010-276913), by the Ministry of Science and Technological 
Development of the Republic of Serbia, under projects No. ON171004 and 
ON173052 and by L'OREAL-UNESCO National Fellowship in Serbia.

\bibliographystyle{elsarticle-num}

\end{document}